\documentclass[conference]{IEEEtran}
\IEEEoverridecommandlockouts
\usepackage{cite}
\usepackage{amsmath,amssymb,amsfonts}
\usepackage{algorithmic}
\usepackage{graphicx}
\usepackage{textcomp}
\usepackage{xcolor}
\usepackage[utf8]{inputenc}
\usepackage{float}

\def\BibTeX{{\rm B\kern-.05em{\sc i\kern-.025em b}\kern-.08em
    T\kern-.1667em\lower.7ex\hbox{E}\kern-.125emX}}
    
\begin{document}

\title{Study Of A Hybrid Photovoltaic-Wind Smart Microgrid Using Data Science Approach}
\author{
\IEEEauthorblockN{Josimar Edinson Chire Saire}
\IEEEauthorblockA{Institute of Mathematics and\\
Computer Science (ICMC) \\
University of São Paulo (USP)\\
São Carlos, SP, Brazil\\
jecs89@usp.br}

\and

\IEEEauthorblockN{José Armando Gastelo-Roque}
\IEEEauthorblockA{Microgrids for Sustainability \\
Mechanical Electrical Engineering \\
Pedro Ruiz Gallo National University\\
Lambayeque, Perú\\
jgastelor@unprg.edu.pe}

\and

\IEEEauthorblockN{Franco Canziani}
\IEEEauthorblockA{Universidad Cientifica del Peru \\
Universita Politecnica de Catalunya \\
Lima, Perú\\
Barcelona, Spain\\
franco@waira.pe}

}

\maketitle


\begin{abstract}


In this paper, a smart microgrid implemented in Paracas, Ica, Peru, composed of 6kWp PV + 6kW Wind and that provides electricity to a rural community of 40 families, was studied using  a data science approach. Real data of solar irradiance, wind speed, energy demand, and voltage of the battery bank from 2 periods of operation were studied to find patterns, seasonality, and existing correlations between the analyzed data.

Among the main results are the periodicity of renewable resources and demand, the weekly behavior of electricity demand and how it has progressively increased from an average of 0.7kW in 2019 to 1.2kW in 2021, and how power outages are repeated at certain hours in the morning when resources are low or there is a failure in the battery bank. These analyzed data will be used to improve sizing techniques and provide recommendations for energy management to optimize the performance of smart microgrids.

\end{abstract}


\section{Introduction}


Lack of access to quality electricity is seen as a limit to improve people's opportunities and quality of life. The role of energy and electricity as a key driver of economic and sustainable development is a fact recognized by the world community, as evidenced by the existence of a specific Sustainable Development Goal (SDG7) that includes access to affordable, reliable energy, sustainable and modern for all by 2030 \cite{RIVA2018203}.


However, the problem is not solved simply by increasing the country's electrification ratio by providing electricity that only allows lighting or through welfare policies. The literature emphasizes that access to electricity must provide "quality energy" and be always accompanied and sustained by other enabling activities and services, in order to contribute to a higher level of education, more business opportunities, gender equity and higher income to local level \cite{RIVA2018203,baldwin_electrification_2015,khandker_shahidur_r_welfare_2009,winther_womens_2017}. One way to measure the quality of the energy provided is through the multilevel framework (MTF) developed by the World Bank. The MTF does not only records whether a home is receiving energy service, but whether this service is "usable" from the household's perspective. The MTF classifies energy access into five levels, from Level 1, when only lighting and DC charging is provided for 4 hours, to Level 5 when electricity is provided for very high-power appliances for a minimum of 23 hours per day \cite{mikul_beyond_2015}.

Smart microgrids are technological alternatives that 
allows electrifying the isolated and remote rural communities and providing modern and quality electricity with MTF Level 5. 

Smart microgrids can accelerate electricity access to areas where the central electricity grid cannot reach in the short to medium term. As these microgrids develop, they can be interconnected, creating a decentralized grid that can aggregate loads and generation capacity, while maintaining the ability to operate as isolated systems should the need arise.

Renewable energy microgrids technology have a higher implementation cost (CAPEX) than conventional technologies like diesel generators. Due to this, it is important sizing the microgrids without oversize the system, to allow their implementation in countries with low budget for rural electrification.

Microgrids for rural electrification are isolated and mostly use local energy resources, such as solar, wind or hydroelectric resources, and sometimes are supported by diesel systems to generate electricity.


One of the main objectives of the microgrid is to supply reliable power to the loads in a microgrid domain. This reliability can be measured through indicators such as the frequency of interruptions, the duration of  discontinuity or the power not supplied \cite{warren2002overview}. Achieving this goal becomes complex when a microgrid system is supplied by renewable energy sources such as wind or solar. Variable wind will cause unpredictable power output in the wind turbine system and intermittent solar irradiance will affect the production of photovoltaic systems. Therefore, such variations are propagated by all subsystems and produce an unpredictable and intermittent electricity generation \cite{ahshan2017microgrid}.

A fundamental factor in the reliability of the system is the relationship between generation and demand. Despite the intermittency, the system can be reliable if there is a match between the periods of largest generation and those of maximum demand. However, an isolated microgrid must guarantee the quality of the energy service regardless of the variability of the generation of resources. Therefore, there are two options: 1) Oversize the generation capacity at the time of sizing (which would reduce the profitability of the project) or 2) Modify the relationship between generation and energy demand \cite{quintero2015impact}(as renewable generation cannot be manipulated, this modification must be made on the demand side) \cite{hussain2016resilient}. The second action is called Management Scenarios (MS) and changing customer demand is called Demand Side Management (DSM) \cite{hussain2016impact}  \cite{ahshan2017microgrid}.

DMS is a concept widely used in the conventional electricity market but it could be applied to optimize and guarantee the reliability of microgrids for rural electrification if seasonality patterns and correlations in energy resources and demand can be identified.

In this paper, a data science approach is used to analyze energy resource data: solar irradiance and wind speed and energy demand for a smart microgrid in Paracas, Ica, Peru as a case study. The smart microgrid studied is made up of a 6kWp photovoltaic system, two 3kW wind turbines and a 38.4kWh lead-acid battery energy storage system that provides electricity to about 40 families. The correlation between demand and resource, increase in demand, periodicity, number of power outages, etc. was studied.

The data analysis will serve as input for a future Demand Side Management (DSM) work on low-cost smart microgrids.




\section{Smart Microgrid Under Study}

\begin{figure*}[hbpt]
\centerline{\includegraphics [width=0.98\textwidth]{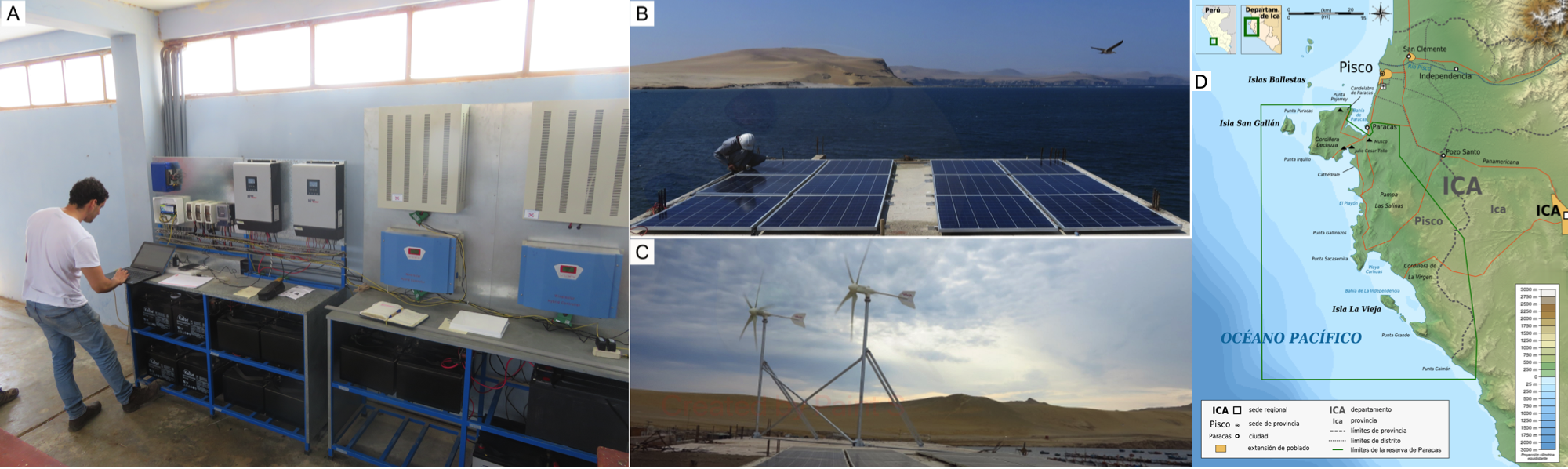}}
\caption{Laguna Grande Smart Microgrid. a) Battery bank and power electronics, b) Photovoltaic System (partial view), c) Wind Turbines, d) Location of the microgrid under study}
\label{fig:avg_ch9_2018}
\end{figure*}

\begin{figure}[hbpt]
\centerline{\includegraphics [width=0.50\textwidth]{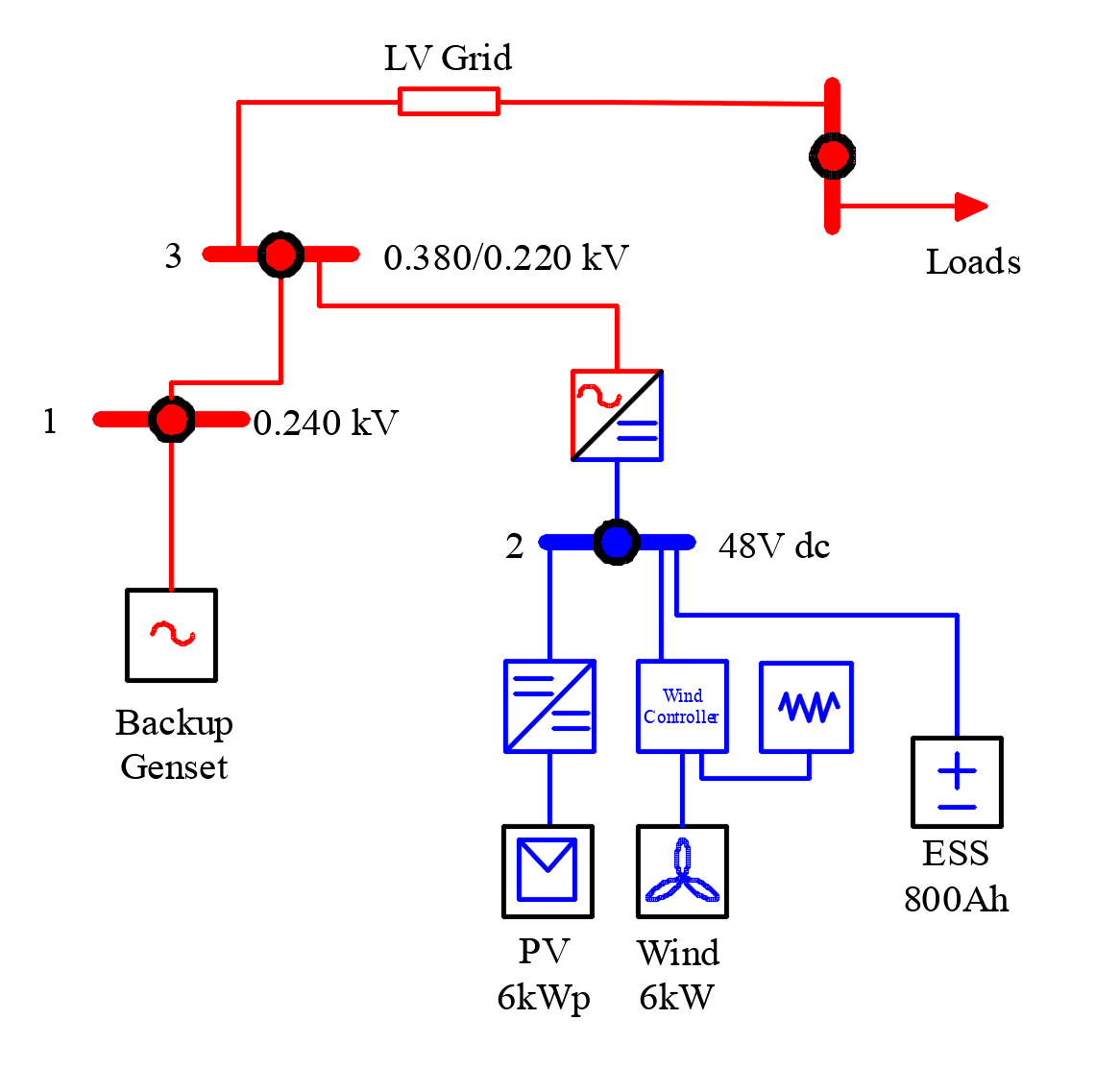}}
\caption{Microgrid configuration one-line}
\label{fig:fftsolar}
\end{figure}
The microgrid under study is located in Laguna Grande, a rural fishing community founded in 1979, located in the Paracas National Reserve, in the Ica region in the coast of Peru. On coordinates 14°08'33.5'' S 76°15'43.6'' W. shows in Fig. 1. Currently, the entire population lives and works around fishing and related activities, and there are almost 90 residents in 35 homes.

The microgrid is composed by PV, energy storage, and wind turbines all connected to a 48 Vdc bus bar and two 48Vdc 4kW inverter/chargers dispatch 230VAC to power all the 35 loads.
Each 3 kW wind turbine has its controller and dump load, and each 3 kWp solar array was assigned to an independent MPPT controller built into the inverter/charger unit. A one-line diagram  of the studied smart microgrid is showed in the Fig. 2. Fig. 1 shows pictures of the microgrid \cite{canziani_hybrid_2021}.

\section{Data and Methods}
The present paper performs experiments with source real data that was collected from the smart microgrid in operation. The following data was measured and collected: Solar Irradiance (W/m\textsuperscript{2}), Wind speed (m/s), Power demand (kW) and DC voltage of the battery bank (V).

Data from two different periods was collected. The first period covers data from December 2018 to June 2019. The second period covers data from October 2020 to February 2021.

\subsection{Solar Irradiance}
The studied smart microgrid use photovoltaic as one of the main source to generate electricity. Solar irradiance was measured as global horizontal irradiance (GHI) and recorded on-site with a Symphonie LI-COR LI-200/R-BL pyranometer.

\subsection{Wind Speed}
The community under study has great wind potential due to its geographical location. Wind direction and speed were recorded at the site with a NRG 200P wind vane and a NRG \#40C anemometer, respectively. All data was stored in a SympohniePlus 3 data logger.

\subsection{Energy Demand and DC Voltage of Battery Bank}
 To measure power demand and Battery Voltage was used AC and DC current meters, and voltage meters connected to an eGauge Pro datalogger.
 
 DC voltage was measured to obtain the times that the system suffered a power outage. DC Voltage has a minimum work value of 43V, if the battery discharges to this value the system turn-off to avoid damage to the battery bank.  

\subsection{Data Mining Process}

The traditional approach for Data Mining includes the next steps: data understanding, pre-processing data, visualization. In the next subsections are explained pre-processing and visualization.

\subsubsection{Pre-processing data}

The exploration of the data shows the inconsistencies in the data, i.e. null values, char values in integer fields. The Language Programming to handle data is Python and packages to process arrays (numpy), handle Dataframes (pandas) and plot (matplotlib).

Besides, the filtering process to select an specific field, i.e. year. It is possible to create a new variable with a substring of date. 

\subsubsection{Processing data and Visualization}

In this context, processing data from a Microgrid. After previous exploration a set of question were purposed. The answer of these questions is composed of creating variables, filtering and select data to plot properly and facilitate the understanding of the answer. The following topics were studied:
\begin{itemize}
\item Typical load curve, battery discharge curve and curves of available resources were calculated. Days with normal working and days with power outages were compared. The typical hours with more frequency of power outages and hour with most available resources were calculated.
\item The existence of seasonality in energy resources and load was studied. An seasonality adjust was applied to the load to analyze if there is a no-seasonal tendency during the 4 year of operation (2018-2021).
\item Solar irradiance, wind speed and power demand were analized using signal processing techniques. If one signal contains a pattern, which repeats itself after a specific period of time, we call it an periodic signal.The time it takes for an periodic signal to repeat itself is called the period P (and the distance it travels in this period is called the wavelength). It is intended to find patterns in this data which can be used to forecast renewable resources and demand using more straightforward forecasting methods.
\item Anomalies like load peaks or atypical solar irradiance o wind speed were searched. And the existence of weekly, monthly or yearly periodicity was studied.  
\item Correlation between solar irradiance and wind speed and between resources and DC voltage of battery bank were studied. 
\item The possibility to made forecasting of solar irradiance, wind speed and load was studied with the objetive of predicting power outages in the microgrid and sent alerts to recommend community turn-off some artifacts to avoid power outages.
        
\end{itemize}

\section{Results}

\begin{figure}[hbpt]
\centerline{\includegraphics [width=0.48\textwidth]{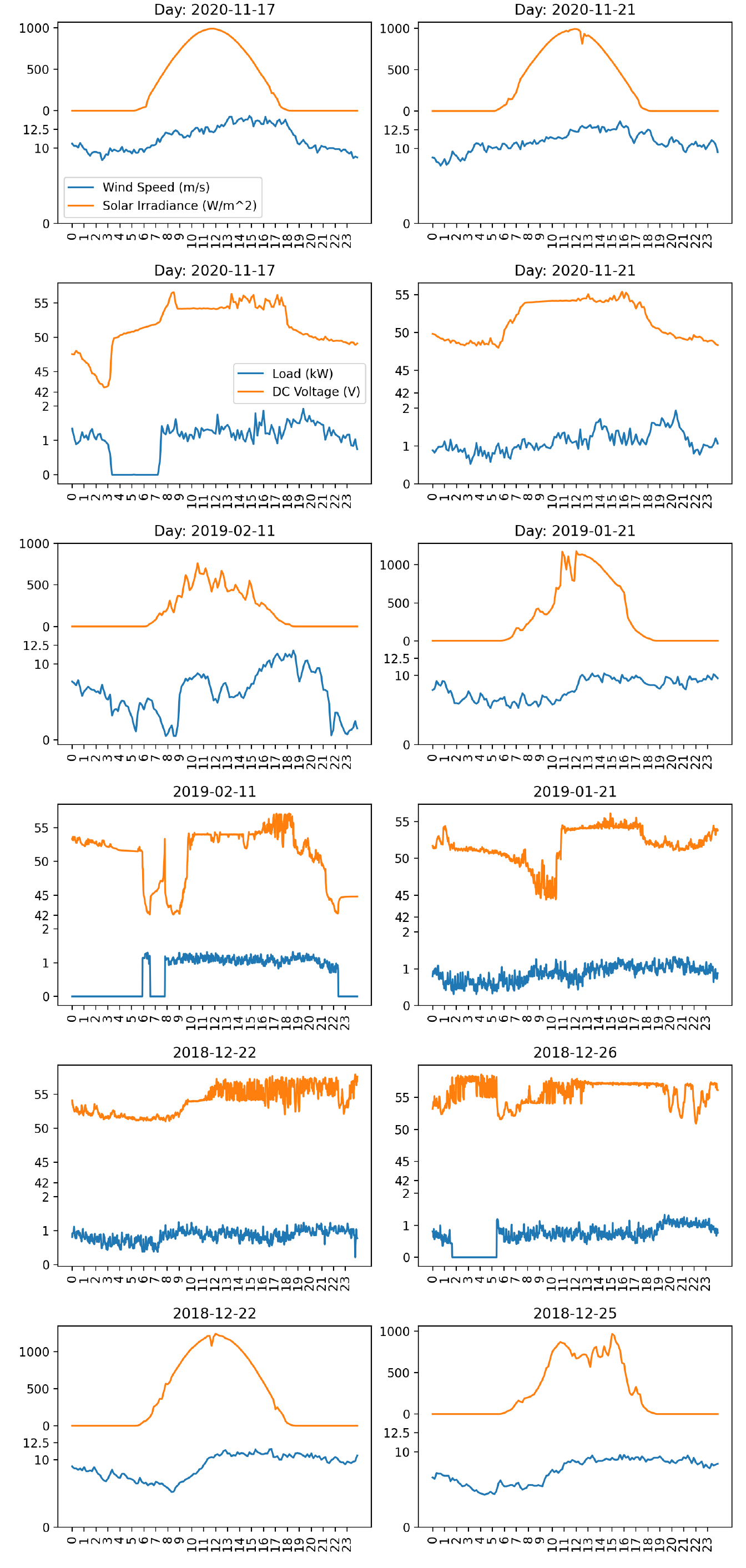}}
\caption{Typical curves for 2018, 2019 and 2020}
\label{fig:avg_up}
\end{figure}

Fig. \ref{fig:avg_up} shows the typical curve of load, resources, and DC voltage for six typical days. Two days for each year are shown, one in which the system worked correctly and another in which there were power outages throughout the day.

\begin{figure}[hbpt]
\centerline{\includegraphics [width=0.45\textwidth]{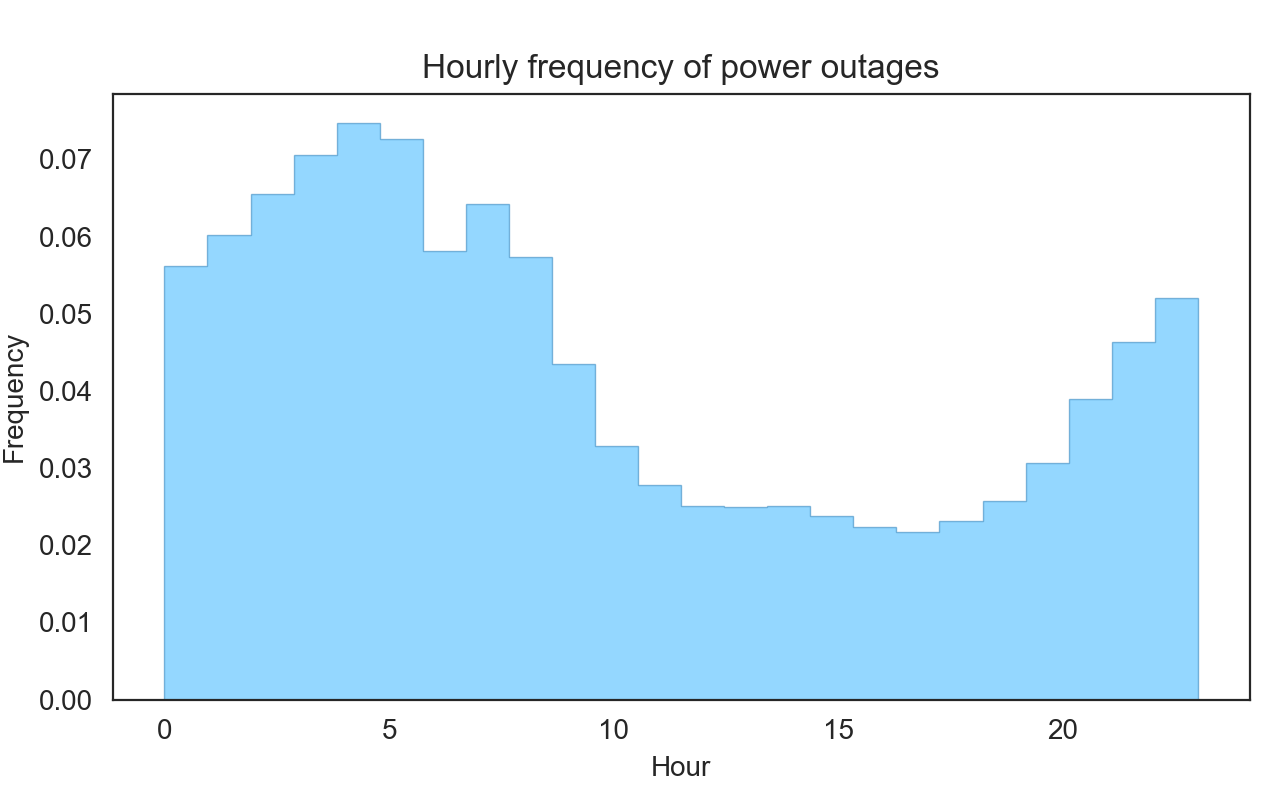}}
\caption{Microgrid configuration one-line}
\label{fig:freqpowerout}
\end{figure}

The hours with more frequency of power outages are in the early morning and in the morning hours, most frequently from 03:00 am to 09:00 am. These hours are precisely those when there is less availability of resources. Fig. \ref{fig:freqpowerout}, shows a histogram of the hours in which a power outages occurred. In total, during the periods analyzed, the microgrid was unable to provide power for around 20 percent of the operating time.

As indicated above, power outages typically occur in the morning. These power outages in the morning hours are due to any of the following reasons: 1) low resource on the previous day or days, 2) high demand prior to power outages, 3) some failure in a battery of the battery bank, 4) the combination of the previous reasons. The data analysis carried out allowed us to obtain the reason for the failures shown in Fig. \ref{fig:avg_up}.

The results show that in 2020-11-17 there was a failure due to the fact that the wind resource during the previous week was very low, after 2020-11-17 there was an increase in wind resource which allowed the normal working of 2020-11-21, despite the existence of the same consumption, this also rules out that there is a fault in the battery bank.

For the cases analyzed in 2019, on 2019-02-11, 2 power outages were recorded throughout the day, one in the morning due to the lack of resources on previous days and one after 10 p.m. The second was due toe both the wind and solar resources were not enough during the day. The solar resource at no time exceeded 700W/m2 and the wind speed was highly variable, with drops close to 1m/s at various times of the day. If we compare the resource of 2019-02-11 with that of 2019-01-21, in which there were no power outages, the difference in the availability of the resource can be noticed. On 2019-02-11, abrupt variations were recorded in the DC Voltage of the battery bank, which could indicate the failure of a battery in the bank.

For the year 2018, a case in which there were power outages and another in which the system worked correctly was also analyzed. The figure shows the operation for the day 2018-12-16 where there was a power outage from 2:00 am to 5:30 am, the resource that was the previous day 2018-12-15 is shown, where it is shown that the wind resource it was a little lower than average and the solar irradiance was highly variable and poor. The lack of recourse the day before produced the failure of 2018-12-16. The performance for 2018-12-22 is shown, where the available resources were optimal and the system worked correctly.

\begin{figure*}[hbpt]
\centerline{\includegraphics [width=0.95\textwidth]{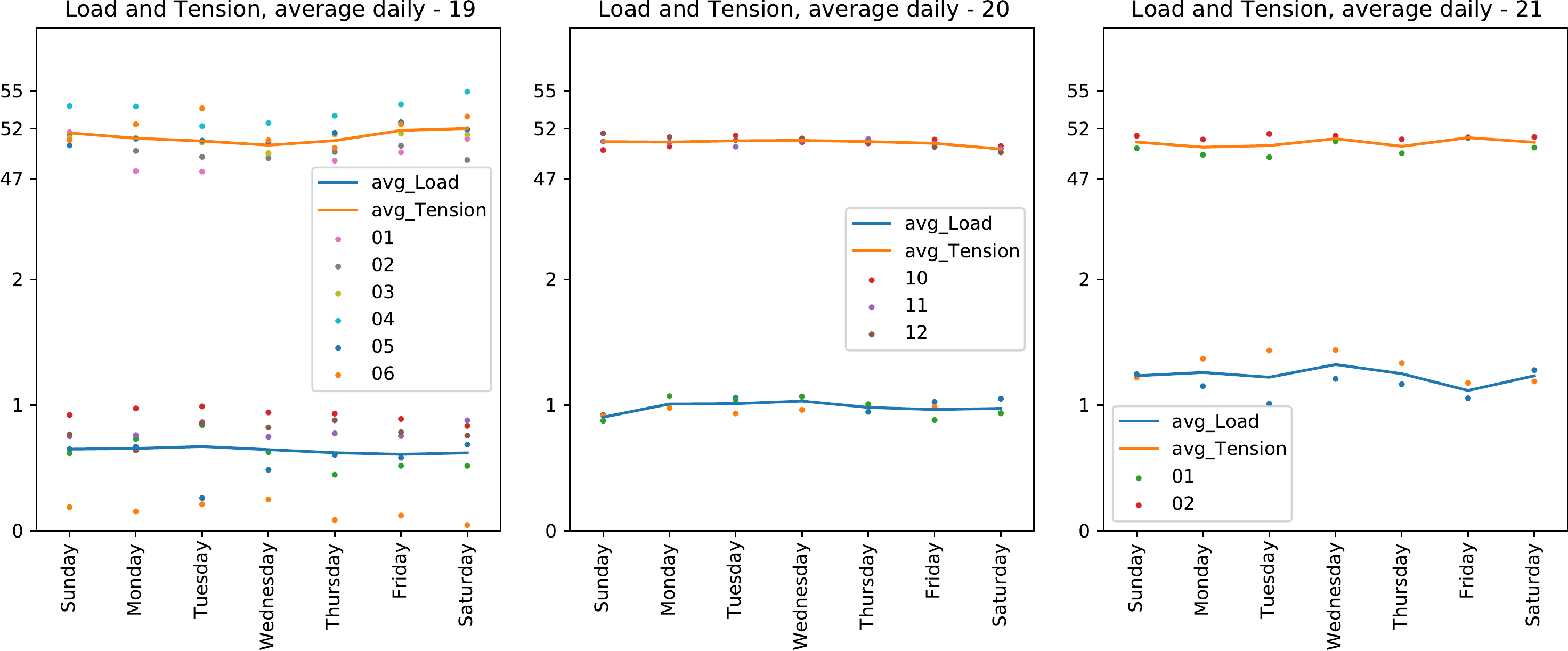}}
\caption{Average Load (kW) and DC Voltage (V)}
\label{fig:avg_use}
\end{figure*}

There are not enough data to establish whether there is annual seasonality in the resource data, the periods analyzed correspond to similar seasons in the years 2018-2019 and 2020-2021. However, the data shows an increase in electricity demand from a maximum average of 0.7kW in 2019 to 1.2kW in 2021 as is shown in Fig. \ref{fig:avg_use}.

 From the analysis through signal processing techniques, the Fast Fourier Transform (FFT) algorithm, the Power Spectral Density (PSD), and the auto-correlation function were applied to solar irradiance, wind speed and electricity demand. From the results, solar irradiance, wind speed and electricity demand present a periodicity every 12 and 24 hours, which was to be expected due to the nature of the solar movement that affects solar irradiance and wind speed and due to because the use of electricity tends to follow a daily cycle. Fig. \ref{fig:fftsolar} show the FFT results.

\begin{figure}[hbpt]
\centerline{\includegraphics [width=0.40\textwidth]{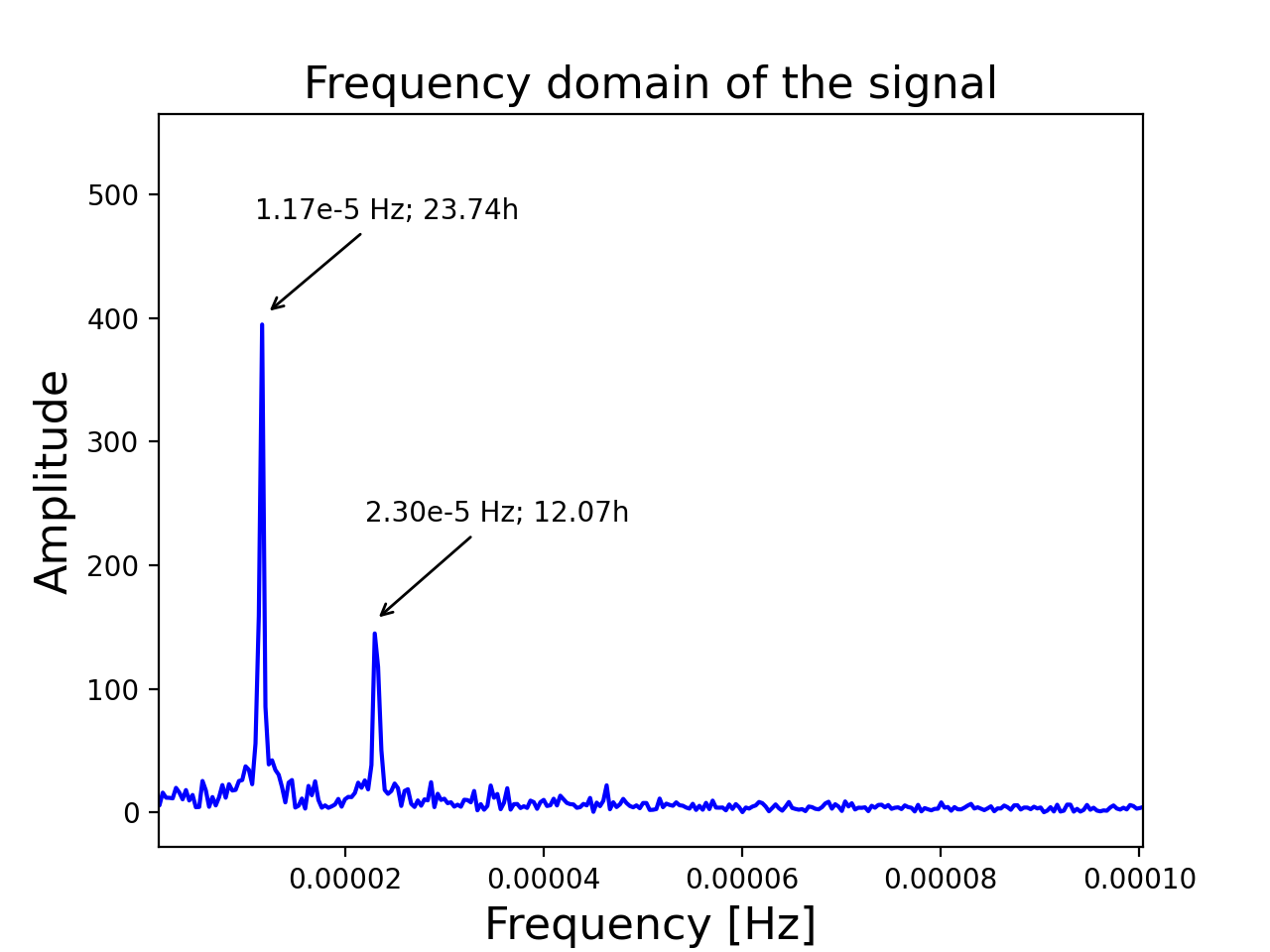}}
\caption{FFT for solar irradiance}
\label{fig:fftsolar}
\end{figure}

There is a clear correlation between solar irradiance and wind speed. The maximum wind speed is reached at the same time as the maximum solar irradiance (around 12 - 1pm), this is a problem for the microgrid since there is a over generation in hours between 11 am to 1 pm that cannot be stored for microgrid use. In addition, in the days that there is low solar resource (700W/m2 are never reached) the wind resource is also low (much less than the average of 10m/s throughout the day).

\section{Conclusions}

In this paper we study the operation of a low power smart microgrid using a data science approach. The data analysis allowed to obtain the following conclusions:
\begin{itemize}
\item There was an increase in the electricity demand of the microgrid by 70 percent from 2019 to 2021, which reflects that access to electricity has allowed the use of additional electrical appliances and therefore an improvement in the community's economy.
\item The microgrid has undergone power outages for 20 percent of the operating time, with a higher frequency of occurrence in the morning hours (3:00 to 9:00 am).
\item The causes of the failures are the existence of an unusable overgeneration at hours of maximum solar and wind resource (11:00 am to 1:00 pm) and the correlation between solar irradiance and wind speed.
\item Solar irradiance and wind speed can be modeled as a periodic signal with periods of 12h and 24h, which allows for simple forecasting in a future job.
\item The Data Science approach is a useful tool that allows knowing different behaviors in microgrids and their causes, for example the cause of power outages in the microgrid by analyzing data prior to the event.
\item These data will be used for future analysis, microgrid improvements, and as input for a future work on Demand Side Management (DSM) of low-cost smart microgrids.
\end{itemize}
\bibliographystyle{IEEEtran}

\bibliography{biblio}

\vspace{12pt}
\end{document}